\documentstyle[12pt,epsfig]{article}
\advance \topmargin by -\headheight
\advance \topmargin by -\headsep
\evensidemargin \oddsidemargin
\begin{document}
\pagestyle{plain}
\begin{titlepage}
\vspace*{0.15cm}

\vspace*{0.5cm}

\hfill {\bf IHEP 2004-12}

\vspace*{2.cm}

\begin{center}
{\Large\bf
   High statistic measurement of the $K^{-} \rightarrow \pi^{0} e^{-} \nu $ 
    decay form-factors} \\

\vspace*{0.15cm}
\vspace*{1.3cm}

{\bf   O.P.~Yushchenko, S.A.~Akimenko, 
 G.I.~Britvich,  K.V.Datsko,  A.P.~Filin, 
A.V.~Inyakin,  A.S.~Konstantinov, V.F.~Konstantinov,  
I.Y.~Korolkov, V.A.~Khmelnikov, V.M.~Leontiev, V.P.~Novikov,
V.F.~Obraztsov,  V.A.~Polyakov, V.I.~Romanovsky, V.M.~Ronjin, 
   V.I.~Shelikhov, N.E.~Smirnov\footnote{Now at INFN, Padova, Italy},
  O.G.~Tchikilev, V.A.Uvarov. }
\vskip 0.15cm
{\large\bf $Institute~for~High~Energy~Physics,~Protvino,~Russia$}
\vskip 0.35cm
{\bf V.N.~Bolotov, S.V.~Laptev,  A.Yu.~Polyarush. }
\vskip 0.15cm
{\large\bf $Institute~for~Nuclear~Research,~Moscow,~Russia$}
\end{center}

\vspace*{3cm}

\begin{abstract}
 The  decay $K^{-} \rightarrow \pi^{0} e^- \nu$ is
 studied using in-flight decays detected with the "ISTRA+" spectrometer. 
 About 920K events are collected for the analysis.  
 The $\lambda_{+}$  slope
 parameter of the decay form-factor  $f_{+}(t)$ in the linear approximation
 (average slope) is measured: 
 $\lambda_{+}^{\mbox{\scriptsize lin}}= 0.02774 \pm 0.00047$(stat) 
 $\pm 0.00032$(syst). 
 The quadratic contribution to the form-factor was estimated to be
 $\lambda^{'}_{+} = 0.00084\pm 0.00027$(stat) $\pm 0.00031$(syst). The 
 linear slope, which has a meaning of $df_+(t)/dt|_{t=0}$ for this fit,
 is $\lambda_{+}= 0.02324 \pm 0.00152$(stat) $\pm 0.00032$(syst).  
  The limits on
 possible tensor and scalar couplings are derived:
 $f_{T}/f_{+}(0)=-0.012 \pm 0.021$(stat) $\pm 0.011$(syst), 
 $f_{S}/f_{+}(0)=-0.0037^{+0.0066}_{-0.0056}$(stat) $\pm 0.0041$(syst).

 \end{abstract}

\end{titlepage}

\newpage
\thispagestyle{empty}

~

\setcounter{page}{0}
\newpage
\raggedbottom
\sloppy

\section{ Introduction}

 The decay  $K \rightarrow e \nu \pi$ ($K_{e3}$) provides unique
 information about the dynamics of the strong interactions. It has been a
 testing ground for such theories as current algebra, PCAC, Chiral Perturbation
 Theory(ChPT).  The study of this decay has a
 particular interest in view of new two-loop order ($\mbox{O}(p^6)$) 
 calculations for K$_{l 3}$ decays in ChPT \cite{Post,Bijnens}.

%
 
 The high-order ChPT calculations
 make a definite prediction for the quadratic term in
 the vector $(f_+(t))$ form-factor and link the scalar  $(f_0(t))$ form-factor 
 linear and quadratic slopes to the  $f_+(0)$ corrections. In turn,  $f_+(0)$ 
 is known to be crucial for the $|V_{us}|$ measurements.
 In fact, 
 the latest measurements do not 
 report any visible non-linearity in the form-factors 
 \cite{KEK1,KEK2,papere,paperm}.
  
In this paper we present a high-statistics measurement
 ($\sim$~919K events) of the Dalitz plot density in this decay. 
 The description of the experimental setup, trigger and reconstruction procedure
 can be found in our previous paper \cite{papere}.

\section{Selection procedure}

 The current analysis is based on the high-statistics data collected 
 during run in Winter 2001. In total, 332M events were logged on tapes.
This statistics is complemented by about 160M MC events generated with 
Geant3 \cite{geant} Monte Carlo program. The MC generation
includes a realistic description of the setup with decay volume 
entrance windows,
tracking chambers windows, chambers gas mixtures, sense wires and cathode 
structures,
\v{C}erenkov counters mirrors and gas, the shower generation in EM calorimeters, 
etc.
 
 The events with one charged track identified as electron and two or three 
additional 
showers in the electromagnetic calorimeter are selected for further processing. 

It was observed, that the main background contribution (about 95\%\ of the
background events) is related to the decay $K^-\to \pi^-\pi^0$, when the
hadronic interaction of the charged pion simulates the electromagnetic
shower in the calorimeter.
Following the method of angular selection used in our analysis of the 
$K_{\mu 3}$ decay \cite{paperm}, we choose the angle between the beam particle
direction and the vector sum of momenta of the final state track and photons as
the variable to perform signal-background separation.

  The expected distribution over this angle is shown in Fig.1 together with the
 real data. One can observe a clear background peak at small values and a good
 agreement of the data and Monte-Carlo. The cut value was selected
 in the region where the maximum of the function 
 $\mbox{Eff}_{\mbox{\scriptsize Signal}}/\mbox{Eff}_{\mbox{\scriptsize Back.}}$ is reached. 

 After that, the momenta of the final state particles were refined with 2C
  $K \rightarrow e \nu \pi^{0}$ kinematic fit. Only the convergence of the fit
  was required. The missing energy  $E_{\nu}=E_{K}-E_{\mu}-E_{\pi^{0}}$
  after 2C fit is shown in Fig.2. This variable
  should exhibit a strong peak at $E_{\nu}\sim 0$ in the presence of noticeable
  background contributions.  The complete absence of any enhancement at small
  values of the missing energy proofs the quality of the selection
  procedure.
   
\begin{minipage}[t]{8.cm}
\epsfig{file=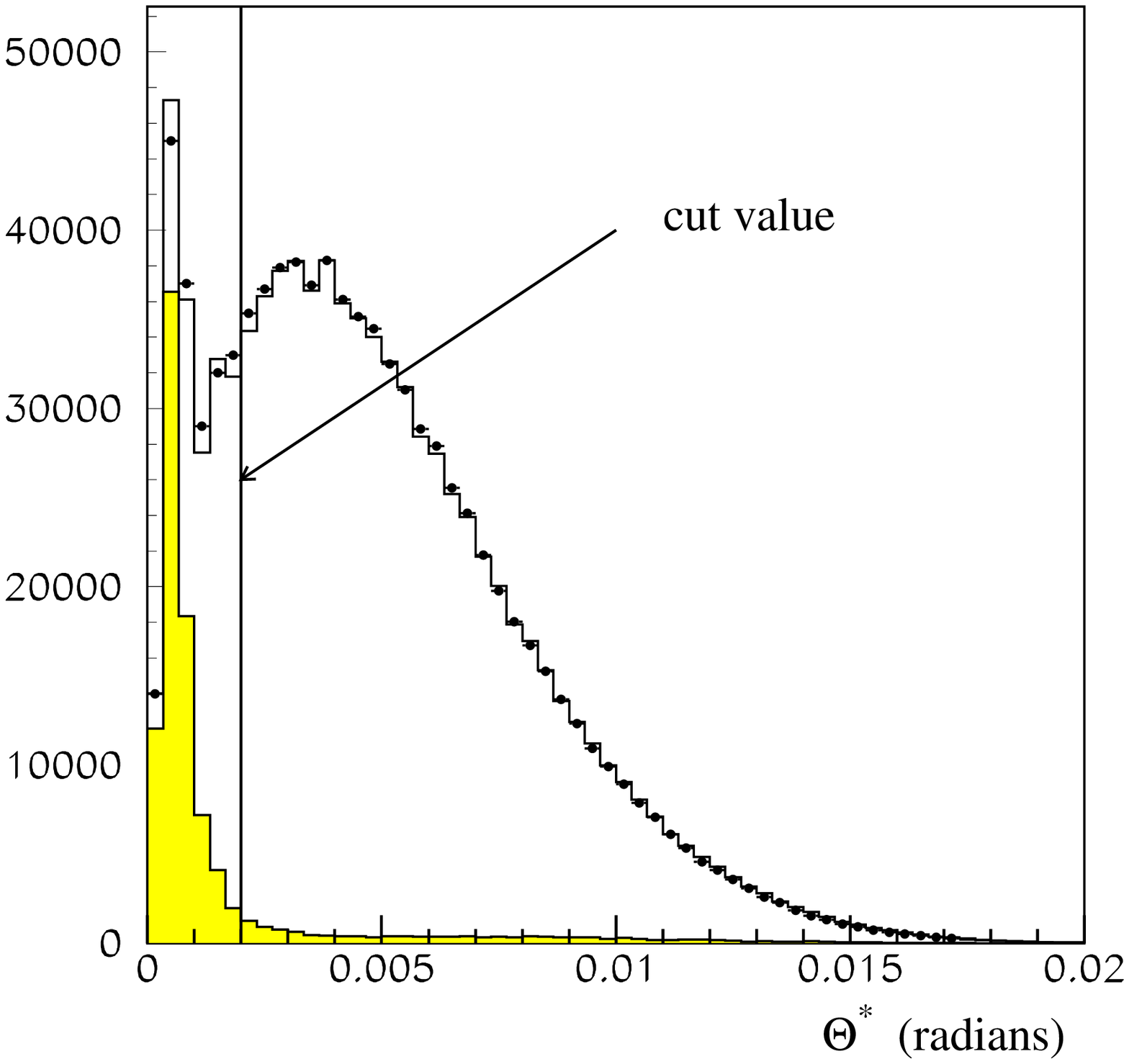,width=8cm}
\begin{center}
Figure 1: Angle between the beam track and the vector sum of final state
particles momenta. The points with errors are data and the solid histogram is 
MC. The shaded area shows the background contribution. 
\end{center}
\end{minipage} \ \hfill \ 
\begin{minipage}[t]{8.cm}
\epsfig{file=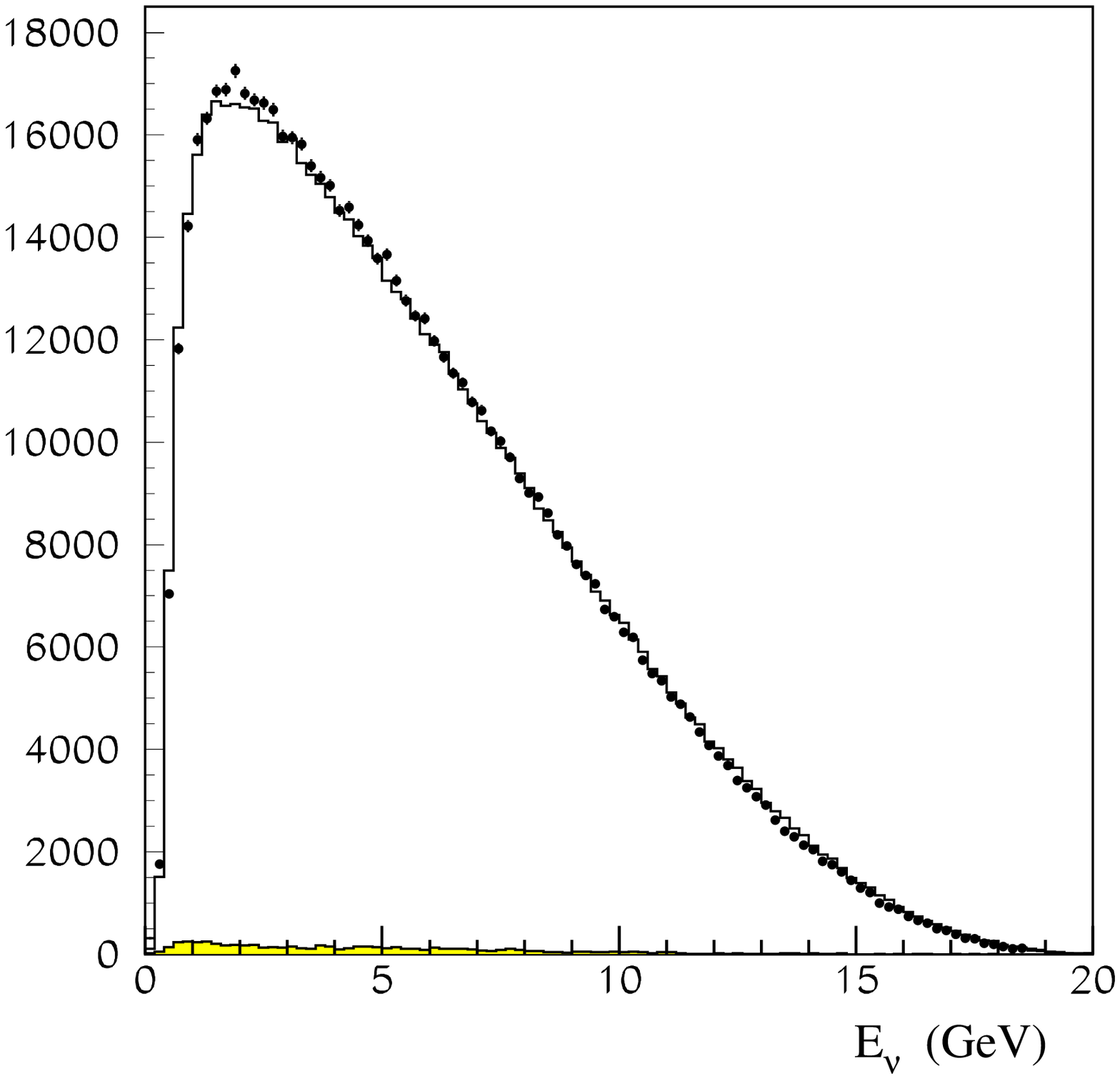,width=8cm}
\begin{center}
Figure 2: Energy of the neutrino compared with MC. The shaded area shows the 
surviving background contribution.
\end{center}
\end{minipage}
 
~

~

The signal Monte-Carlo events for Figures 1 and 2 are weighted with the 
$K_{e 3}$ matrix element where we use  $\lambda_{+} = 0.0286$ (fixed from 
our preliminary measurements \cite{papere}). 

 Finally, 919K events were selected. 
 We estimate the 
 surviving background contribution to be around 2.1\%.

\section{ Analysis}

The most general Lorentz-invariant form of the matrix element for the 
$K^{-} \rightarrow l^{-} \nu \pi^{0}$ decay is \cite{Steiner}:
\begin{equation}
M= \frac{G_{F}V_{us}}{2} \bar u(p_{\nu}) (1+ \gamma^{5})
[2m_{K}f_{S} -
[(P_{K}+P_{\pi})_{\alpha}f_{+}+
(P_{K}-P_{\pi})_{\alpha}f_{-}]\gamma^{\alpha} + i \frac{2f_{T}}{m_{K}}
\sigma_{\alpha \beta}P^{\alpha}_{K}P^{\beta}_{\pi}]v(p_{l})
\end{equation}
It consists of scalar, vector, and tensor terms. The $f_{\pm}$ form-factors
are the functions of $t= (P_{K}-P_{\pi})^{2}$. In the Standard Model (SM),
the W-boson exchange leads to the pure vector term. 
The scalar and/or tensor terms which are ``induced'' by
 EW radiative corrections are negligibly
small, i.e  nonzero scalar or tensor form-factors would indicate 
the physics beyond the SM. 

The term in the vector part, proportional to $f_{-}$, 
is reduced (using the Dirac
equation) to the scalar form-factor. In the same way, the tensor term is 
reduced to
a mixture of the scalar and  vector form-factors. The redefined vector (V) and 
scalar (S) terms, and the corresponding Dalitz plot
density in the kaon rest frame ($\rho(E_{\pi},E_{l})$) are \cite{Chizov}:
\begin{eqnarray}
\rho (E_{\pi},E_{l}) & \sim & A \cdot |V|^{2}+B \cdot Re(V^{*}S)+C \cdot |S|^{2} \\
 V & = & f_{+}+(m_{l}/m_{K})f_{T} \nonumber \\ 
S & = & f_{S} +(m_{l}/2m_{K})f_{-}+
\left( 1+\frac{m_{l}^{2}}{2m_{K}^{2}}-\frac{2E_{l}}{m_{K}}
-\frac{E_{\pi}}{m_{K}}\right) f_{T} \nonumber \\ 
A & = & m_{K}(2E_{l}E_{\nu}-m_{K} \Delta E_{\pi})-  
m_{l}^{2}(E_{\nu}-\frac{1}{4} \Delta E_{\pi}) \nonumber \\
B & = & m_{\l}m_{K}(2E_{\nu}-\Delta E_{\pi}) ;~ E_{\nu}=m_{K}- E_{l}-E_{\pi}
 \nonumber \\
C & = & m_{K}^{2} \Delta E_{\pi};~ \Delta E_{\pi}  =  E_{\pi}^{max}-E_{\pi} ;~
E_{\pi}^{max}= \frac{m_{K}^{2}-m_{l}^{2}+m_{\pi}^{2}}{2m_{K}} \nonumber 
\end{eqnarray}
With the selected number of events we can not neglect the $V-S$ interference
term proportional to the electron mass. 
The term proportional to $f_-$ is neglected in our
 analysis.
 
For further analysis we assume a general quadratic 
dependence of $f_{+}$ on t:
\begin{equation}
 f_{+}(t)=f_{+}(0)\left( 1+\lambda_{+} t/m_{\pi}^2+\lambda_{+}^{'}t^2/m_{\pi}^4 
 \right).
\end{equation}

The procedure of the extraction of the form-factor parameters
starts with the subdivision of the 
  Dalitz plot region 
$ y= 0.12 \div 0.92;\; z=0.55 \div 1.075$ $(y=2E_e/m_K,\; z=2E_{\pi}/m_K)$
 into $100\times 100$ bins.

The signal MC was generated with the constant matrix element and 
 the amplitude-induced weights should be calculated during the fit procedure. 
 One can
observe that 
 the Dalitz-plot density function $\rho (y,z)$ in (2) can be presented in the 
factorisable form: 
\begin{equation}
\rho (y,z)=  \sum_{\alpha}{F_{\alpha}(\lambda _{+},\lambda_{+}^{'}
, f_{S}, f_{T}) \cdot K_{\alpha}(y,z)}, 
\end{equation}
where $F_{\alpha}$ are simple  bilinear functions of the form-factor parameters
 and $K_{\alpha}(y,z)$ are the
kinematic functions which are calculated from the MC-truth information.
For each $\alpha$, the sums of $K_{\alpha}(y,z)$ over events  are accumulated 
in the
Dalitz plot bins (i,j) to which the MC events fall after the reconstruction.
Finally, every bin in the Dalitz plot gets weights $W_{\alpha}(i,j)$ and
the density function $r(i,j)$ which enters into the fitting procedure is 
constructed:
\begin{equation}
r(i,j)= \sum_{\alpha}{F_{\alpha}(\lambda _{+},\lambda_{+}^{'}
, f_{S}, f_{T}) \cdot W_{\alpha}(i,j)}
\end{equation}

 This method allows one to avoid the
systematic errors due to the ``migration'' of the events over the Dalitz plot
due to the finite experimental resolution and automatically takes into account
the efficiency of the reconstruction and selection procedures.

To take into account the finite number of MC events in the particular bin and
strong variation of the real data events over the Dalitz plot, we minimize a 
${-\cal L}$ function defined as \cite{Saclay}:
\begin{equation}
-{\cal L} = 2\sum_j n_j\ln\left[ \frac{n_j}{r_j}\left( 1-\frac{1}{m_j+1}\right)
 \right] + 2\sum_j (n_j+m_j+1)\ln\left(
\frac{1+\frac{r_j}{m_j}}{1+\frac{n_j}{m_j+1}}\right),
\end{equation}
where the sum runs over all populated bins, and $n_j$, $r_j$ and $m_j$ are the
number of data events, expected events and generated Monte Carlo events
respectively. For large $m_j$ Eq. (8) reduces to the more familiar expression
\begin{displaymath}
-{\cal L} = \sum_j [2(r_j-n_j) + 2 n_j\ln n_j/r_j]
\end{displaymath}

 The radiative corrections were taken into account by re-weighting every
 Monte-Carlo event, using MC-truth information,  according to the recent
 calculations in \cite{RC}.

The minimization is performed by means of the ``MINUIT'' 
program \cite{Minuit}.  The errors are calculated by ``MINOS'' procedure of 
``MINUIT'' at the level $\Delta {\cal L} = 1$, corresponding to 68\%\ 
coverage probability for 1 parameter.

\section{Results}

A  fit of the $K_{e3}$ data with 
$f_{S}=f_{T}=\lambda_{+}^{'}=0$ gives $\lambda_{+}=0.02774 \pm 0.00047$.
The total number of
bins is 6991 and $\chi^2/\mbox{ndf} = 0.97$. The quality of the fit is
illustrated in figures 3 and 4 where the projected variables $y=2E_{e}/m_K$
and  $z=2E_{\pi^0}/m_K$ are presented.

\begin{minipage}[t]{8.cm}
\epsfig{file=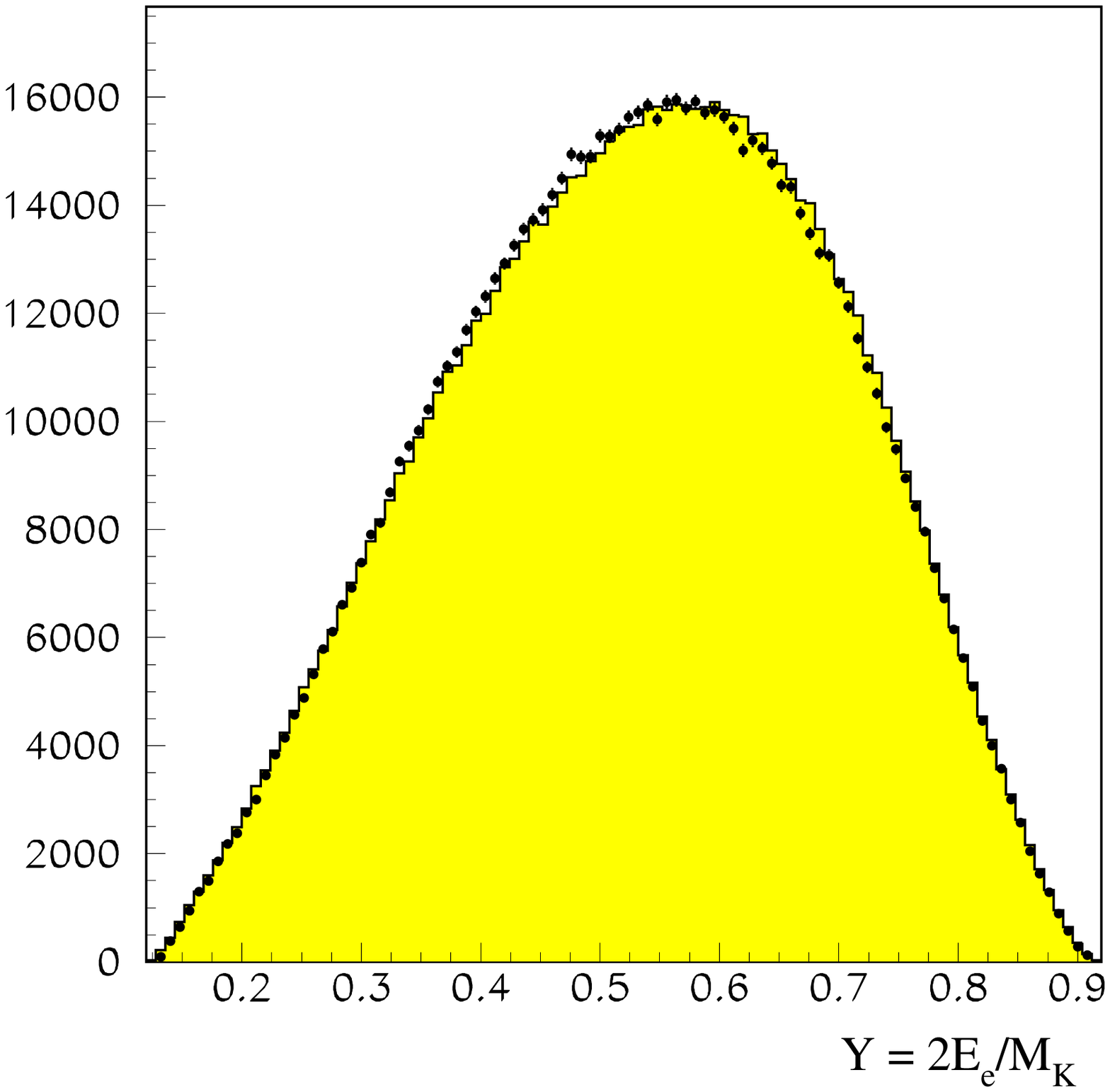,width=8cm}
\begin{center}
Figure 3: Y distribution. \\ The points with errors are the real data \\ and 
the shaded area -- signal MC.
\end{center}
\end{minipage} \ \hfill \ 
\begin{minipage}[t]{8.cm}
\epsfig{file=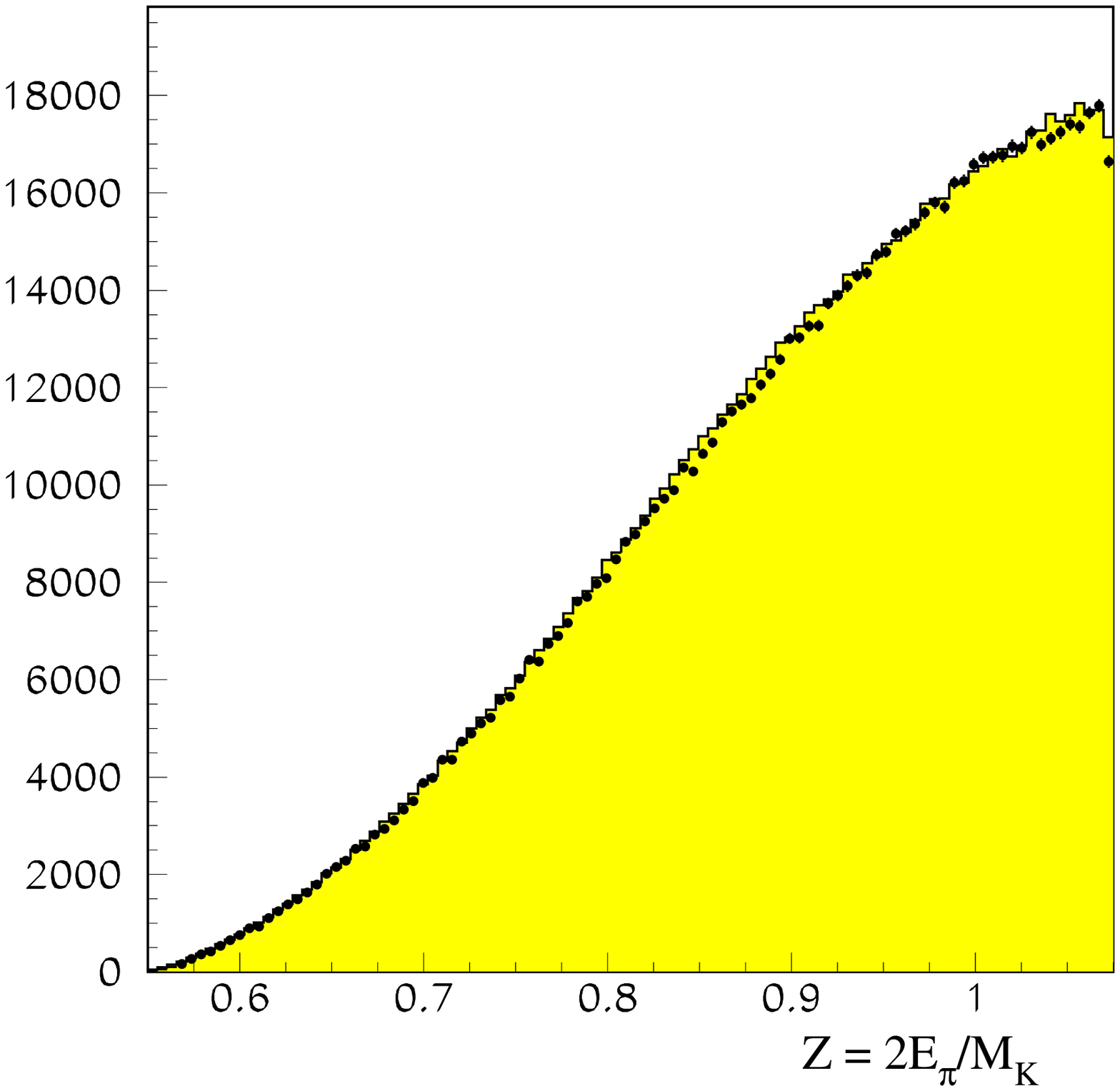,width=8cm}
\begin{center}
Figure 4: Z distribution. \\ The points with errors are the real data \\ and 
the shaded area -- signal MC.
\end{center}
\end{minipage}
 
~

~

The Table 1 represents fits  with possible nonlinear 
term in $f_+$ (Eq. 3) as well as the fits with tensor and
scalar contributions (Eq. 1).

Every row of the Table 1 represents a particular fit where the parameters 
shown without errors are fixed. To qualify the statistical significance of the
particular fit, we also show the change in the $\chi^2$ value obtained 
 with respect to the fit without non-linear or
anomalous contributions.

The second row shows a fit where the nonlinearity is allowed in $f_{+}(t)$.
One can observe 
$\lambda_{+}-\lambda_{+}'$ (Fig.5) correlation that results in a significant
$\lambda_{+}$ errors enhancement and visible shift of $\lambda_{+}$
parameter.

\renewcommand{\arraystretch}{1.4}

\begin{center}
\begin{tabular}{|ccccc|}
\hline
  $\lambda_{+}$ &  $\lambda_{+}^{'}$ & 
   $f_T/f_{+}(0)$ & $f_S/f_{+}(0)$ &  $\Delta\chi^2$ \\ \hline\hline
  $0.02774\pm 0.00047$ & 0. & 0. & 0. & 0. \\ \hline
  $0.02324\pm 0.00152$ & $0.00084\pm 0.00027$ & 0. & 0. & -9.8 \\ \hline
  $0.02774\pm 0.00047$ & 0. & $-0.012\pm 0.021$ & 0. & -0.3 \\ \hline
  $0.02771\pm 0.00047$ & 0. & 0. & $-0.0059^{+0.0089}_{-0.0054}$ & -0.5 \\ \hline
  $0.02324\pm 0.00152$ & $0.00084\pm 0.00027$ &$-0.012\pm 0.021$ & 0. & -9.9 \\
  \hline
  $0.02325\pm 0.00152$ & $0.00084\pm 0.00027$ & 0. & 
   $-0.0037^{+0.0066}_{-0.0056}$  & -9.9 \\ \hline
  \hline
\end{tabular}

\begin{center}
Table 1. The $K_{e 3}$ fits.
\end{center}
\end{center}

\begin{center}
\epsfig{file=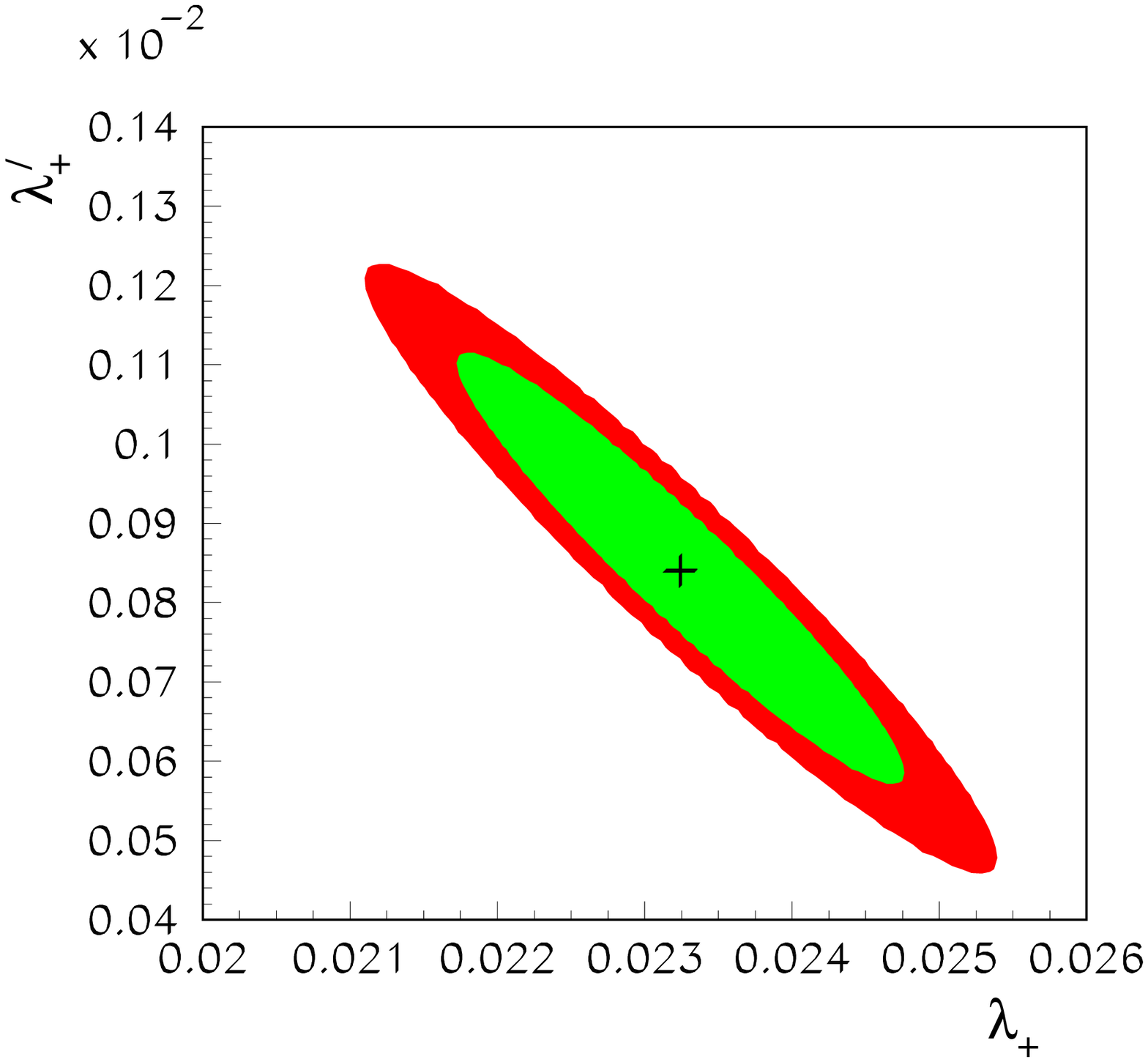,width=10cm}
\end{center}
\begin{center}
Figure 5: $1\sigma$ 
and $2\sigma$ contours in the $\lambda_+ - \lambda^{'}_{+}$ 
plane. 
\end{center}

. We should consider the $\lambda_{+}$ parameter in the
linear approximation as a mean slope over all the physical $q^2$-region, while 
the true value of the linear slope ($df_+(t)/dt|_{t= 0}$) should be taken from
the non-linear fit.

We also perform model-independent one-dimensional ($y$) 
fits where the data in every of the 100 
$q^2/m_{\pi}^2$ bins were fitted independently. 
The resulting distribution is shown in Fig.6.
The normalization $f_+(0) = 1$ is assumed. The visible non-linearity can be
observed in Fig.7, where the ratio $f_+(t)/f_+(0)/(1+\lambda_+ q^2/m_{\pi}^2)$
is presented. The parabolic curve represents the fit with the quadratic
non-linearity in the form-factor.

\begin{minipage}[t]{8.cm}
\epsfig{file=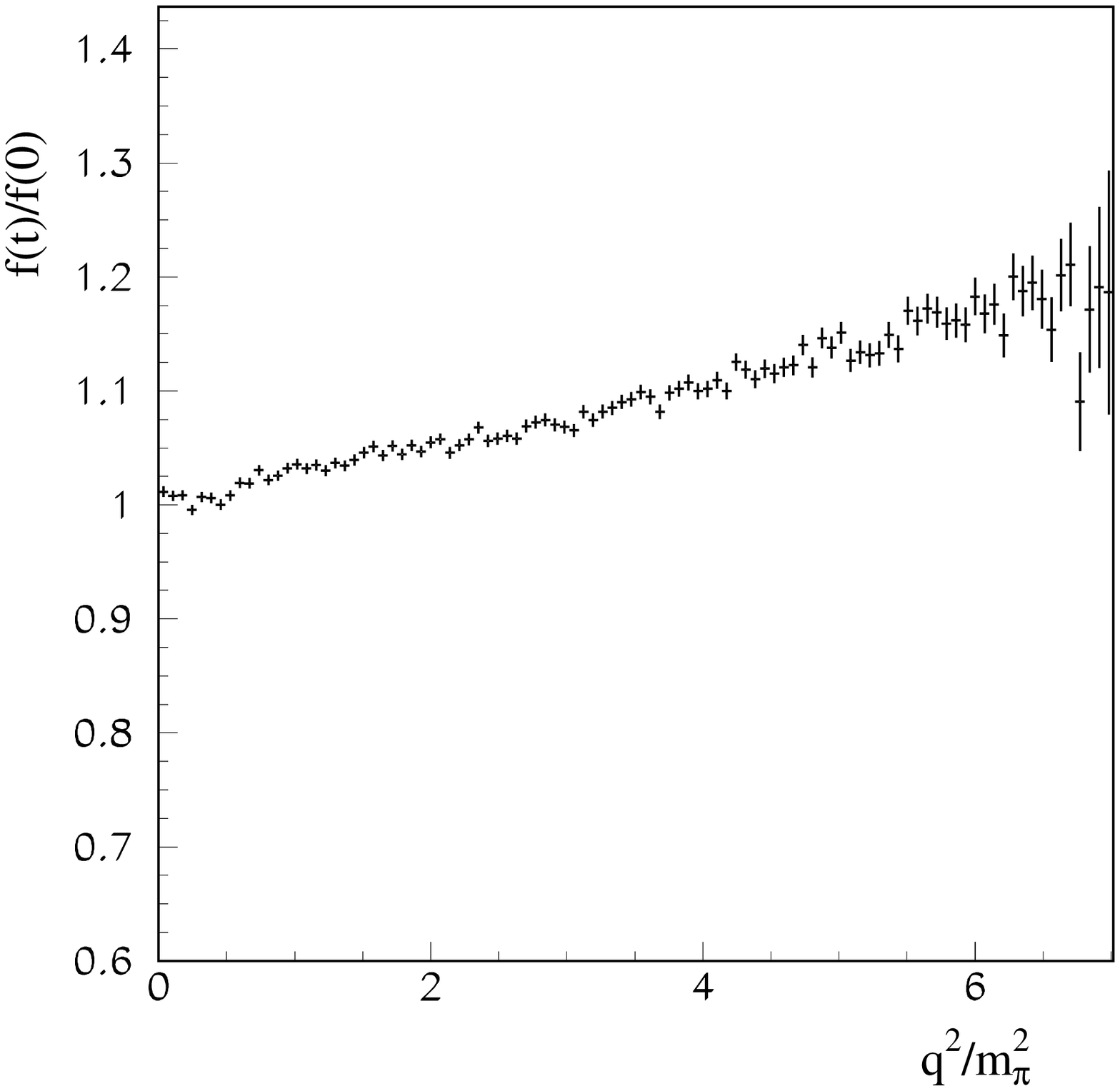,width=8cm}
\begin{center}
Figure 6: The value of $f_+(t)/f_+(0)$ obtained in the model-independent fits.
\end{center}
\end{minipage} \ \hfill \ 
\begin{minipage}[t]{8.cm}
\epsfig{file=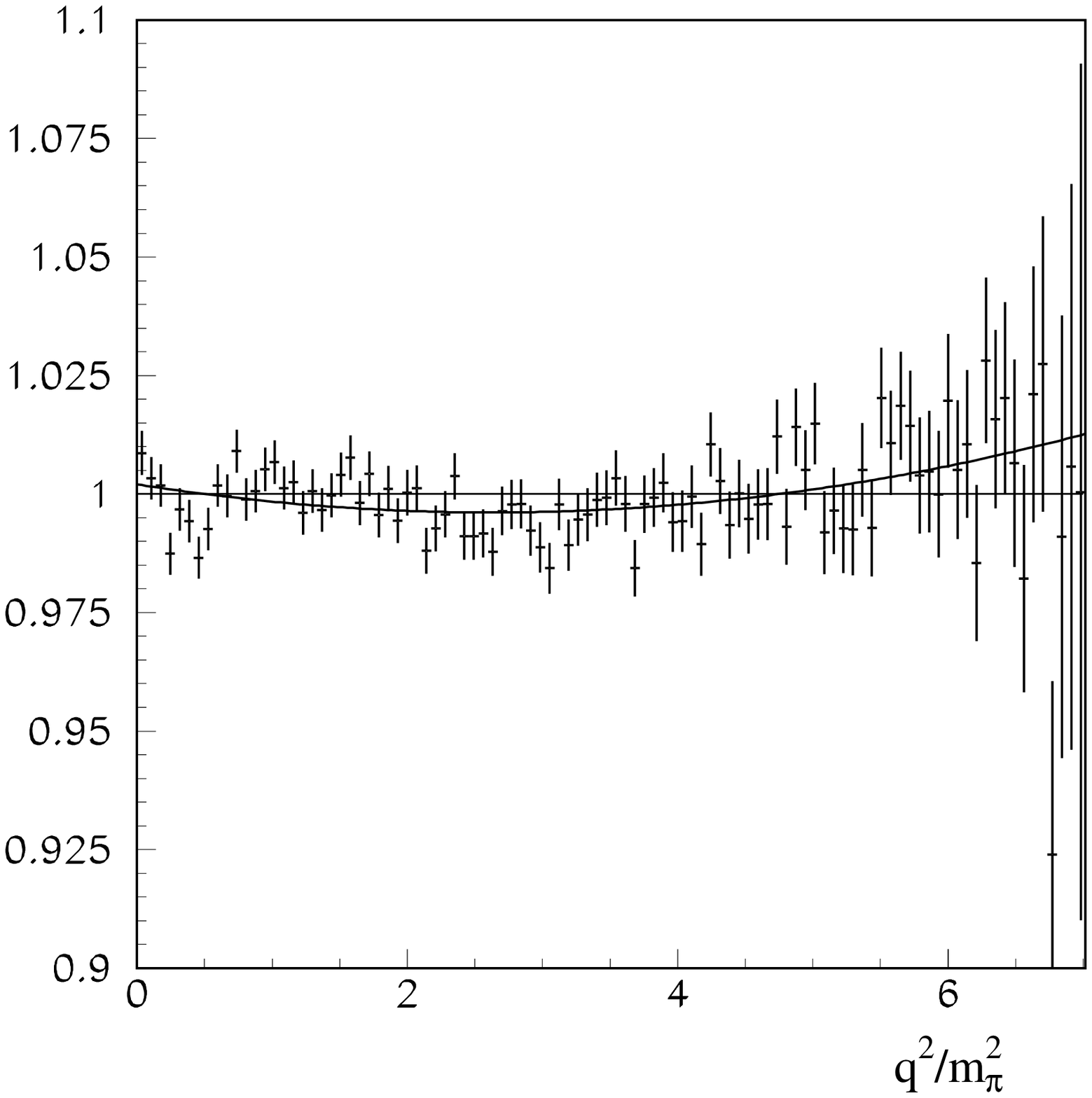,width=8cm}
\begin{center}
Figure 7: The value of $f_+(t)/f_+(0)/(1+\lambda_+ q^2/m_{\pi}^2)$.
The fit with non-linear contribution is shown. 
\end{center}
\end{minipage}
 
~

~

 This non-linearity can not be explained by a possible scalar contribution
 (that also results in the enhancement of the number of events at large values
 of $q^2$). The row 4 of the Table 1 represents a search for the scalar
 term with the vector form-factor set to be linear. 
 The resulting value of $f_{S}/f_+(0)$ is compatible with zero.

 We also perform a model-independent fit to extract simultaneously 
 $f_+(t)$ and $f_S(t)$. The resulting distribution for the value $f_S(t)/f_+(0)$
 is shown in Fig.8. The value of the scalar contribution is compatible with zero
 with strong enhancement of the errors at small values of $t$. This enhancement
 is explained by the dependence of the scalar contributions (Eq. 2) on the 
 Dalitz variables. One can observe that the leading term $|S|^2$ is proportional
 to $t$ and vanishes at $t\to 0$.
 
 The last row of the Table 1 represents a fit with both 
 scalar contribution and the
 quadratic term in the vector form-factor.
 
 We also do not see any tensor contribution in our data (rows 3 and 5 in the 
 Table 1).

~

\begin{center}
\epsfig{file=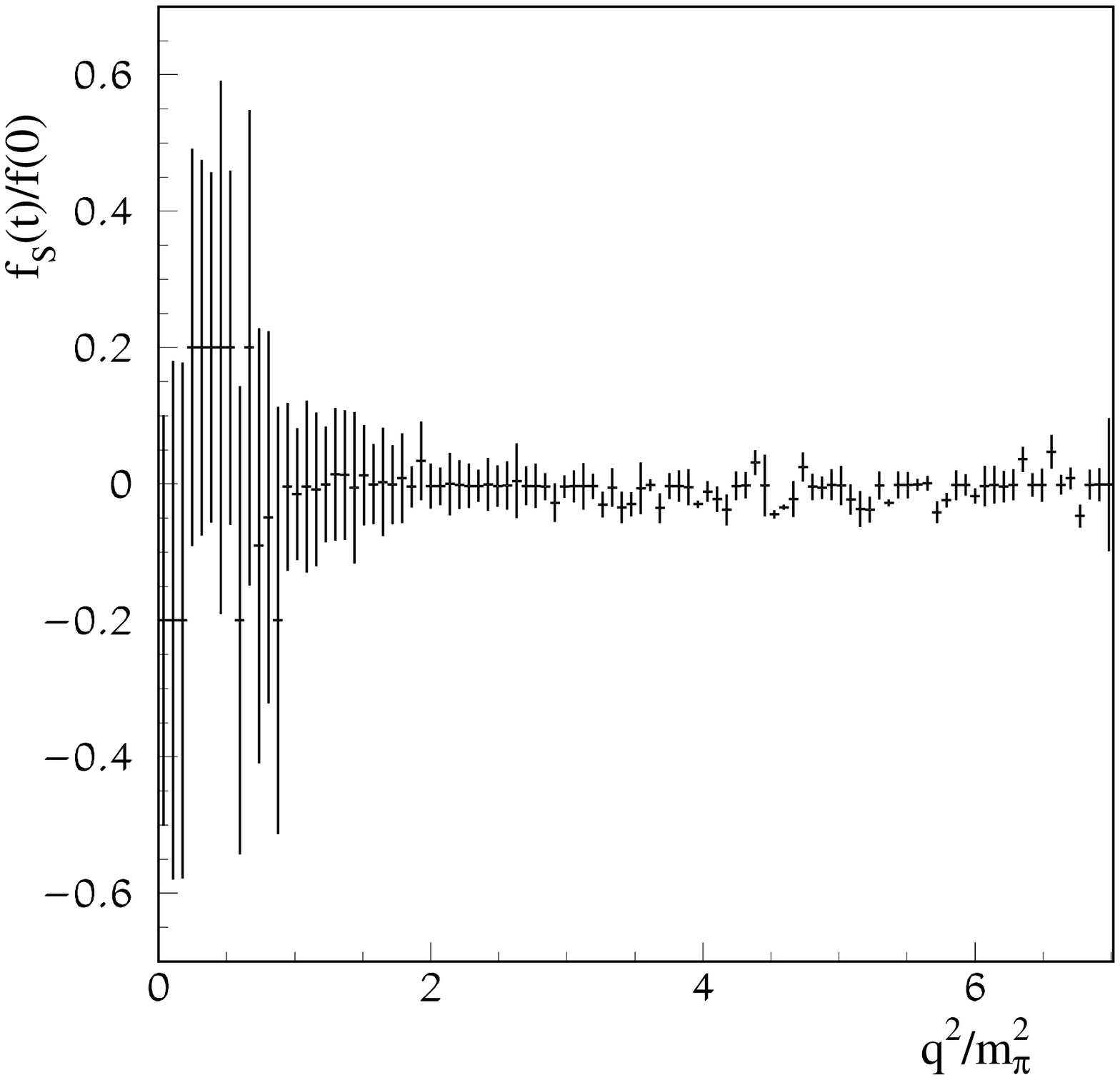,width=10cm}
\end{center}
\begin{center}
Figure 8: The value of $f_S(t)/f_+(0)$ obtained in the model-independent fit.
\end{center}

Different sources of systematics are investigated. 
We allow variations of the electron selection cuts and angular 
cut. 
The Dalitz plot binning, signal and
background MC variations are also applied.

The systematic errors are summarized in the Table 2.
We can conclude that the main contribution into systematic error comes from the
main selection cut variation and from the limited amount of the background MC.

\begin{center}
\begin{tabular}{|c|rrrr|} \hline
Source &   $\lambda_{+}$~~ &  $\lambda_{+}^{'}$~~ & 
   ~~$f_T/f_{+}(0)$ & ~~$f_S/f_{+}(0)$ \\ \hline\hline
 $e^-$ selection     & 0.00017 & 0.00013 & 0.005 & 0.0020 \\
 angular cut         & 0.00020 & 0.00021 & 0.006 & 0.0020 \\
 Dalitz plot binning & 0.00004 & 0.00006 & 0.001 & 0.0005 \\
 Signal MC variation & 0.00006 & 0.00004 & 0.002 & 0.0006 \\
 Backg. MC variation & 0.00016 & 0.00018 & 0.008 & 0.0026 \\ \hline
  Total              & 0.00032 & 0.00031 & 0.011 & 0.0041 \\ \hline
\end{tabular}
\end{center}

\begin{center}
Table 2. The systematic error contributions.
\end{center}

~

\section{Summary and conclusions}
The $K^{-}_{e3}$ decay has been studied using in-flight decays of 25 GeV 
$K^{-}$ detected by the ``ISTRA+'' magnetic spectrometer. 

 The $\lambda_{+}$ parameter of the vector form-factor in the linear
 approximation (average slope)
 is measured to be: 
\begin{center} 
 $\lambda_{+}^{\mbox{\scriptsize lin}}=0.02774 
 \pm 0.00047\; (stat) \pm 0.00032\; (syst).$
\end{center}
\noindent 

A visible non-linear contribution is observed for the first time :
\begin{center} 
 $\lambda_{+}^{'}=0.00084 \pm 0.00027\; (stat) \pm 0.00031\; (syst).$
\end{center}
With the quadratic term in the vector form-factor the linear slope 
(which has a meaning of $df_+(t)/dt|_{t=0}$) is determined to be:
\begin{center} 
 $\lambda_{+}=0.02324 \pm 0.00152\; (stat) \pm 0.00032\; (syst).$
\end{center}
\noindent 
The limits on 
 possible tensor and scalar couplings are derived from the fit with quadratic
 vector form-factor: 
\begin{center} 
 $f_{T}/f_{+}(0)=-0.012 \pm 0.021\; (stat) \pm 0.011\; (syst) ; $ \\[3mm]
 $f_{S}/f_{+}(0)=-0.0037^{+0.0066}_{-0.0056}\; (stat) \pm 0.0041\; (syst).$ 
\end{center}

 The value of the quadratic term in the vector form-factor is in a good
 agreement (within the statistical errors) with the 
 $\mbox{O}(p^6)$ ChPT predictions \cite{Post,Bijnens}.

The obtained limits on the scalar and tensor terms can be used to get 
limits on several
exotic models.

 The leptoquark-induced amplitudes for $K_{e3}$ decay 
 were considered in \cite{lepto1}. In this model
the relation between tensor and
 scalar terms is fixed: 
 $f_T = -6.87 f_S$. The fit of our data with this constraint 
 gives  the following scalar
 contribution: $f_S/f_+(0) = 0.0046^{+0.0042}_{-0.0067}$. This result can be
 converted in the upper limit $|f_S/f_+(0)| < 0.011$ (90\%  C.L.).
Using the expression in \cite{lepto1}:
\begin{equation}
 \frac{f_S}{f_+(0)} = \frac{\sqrt{2}}{16 G_F |V_{su}|} 
 \frac{m_K^2 - m_{\pi}^2}{(m_s - m_u)m_K} \frac{1}{\Lambda^2_{LQ}},
\end{equation}
where $\Lambda_{LQ}$ is the ratio of leptoquark mass to the square of the
Yukawa-like coupling, we get
\begin{displaymath}
 \Lambda_{LQ} > 2.55\mbox{ TeV (90\% C.L.)}
\end{displaymath}

\vspace*{0.5cm}

{\bf Acknowledgments}

We would like to thank Vincenzo Cirigliano for providing us with the 
program for the radiative correction calculations.

\vspace*{0.5cm}

The work
is  supported by the RFBR grant N03-02-16330. \\

\end{document}